\documentclass[12pt]{article}
\usepackage{amsmath}%\pdfoutput=1
\usepackage{color}
\usepackage{graphicx}
\usepackage[footnotesize, skip=-25pt]{caption}

\usepackage{subcaption}

\begin{document}
\begin{flushright}
KANAZAWA-15-06\\
May, 2015
\end{flushright}
\vspace*{1cm}

\begin{center}
{\Large\bf Constrained inflaton due to a complex scalar}
\vspace*{1cm}

{\Large Romy H. S. Budhi}~$^{1,2,}$\footnote[1]{e-mail:
~romyhanang@hep.s.kanazawa-u.ac.jp}, ~
{\Large Shoichi Kashiwase}~$^{2,}$\footnote[2]{e-mail:
  ~shoichi@hep.s.kanazawa-u.ac.jp},\\
{\Large and}\\
{\Large Daijiro Suematsu}~$^{2,}$\footnote[3]{e-mail:
~suematsu@hep.s.kanazawa-u.ac.jp}
\vspace*{0.5cm}\\

$^1${\it Physics Department, Gadjah Mada University, 
Yogyakarta 55281, Indonesia}\\
$^2${\it Institute for Theoretical Physics, Kanazawa University, 
Kanazawa 920-1192, Japan}
\end{center}
\vspace*{1.5cm} 

\noindent
{\Large\bf Abstract}\\
We reexamine inflation due to a constrained inflaton
in the model of a complex scalar. Inflaton evolves along 
a spiral-like valley of special scalar potential in the scalar 
field space just like single field inflation. 
Sub-Planckian inflaton can induce sufficient $e$-foldings because of
a long slow-roll path. In a special limit, 
the scalar spectral index and the tensor-to-scalar ratio has 
equivalent expressions to the inflation with monomial potential 
$\varphi^n$. The favorable values for them could be obtained 
by varying parameters in the potential. 
This model could be embedded in a certain radiative neutrino 
mass model.  
\newpage
%%%%%%%%%%%%%%%%%%%%%%%%%%%%%%%%%%%

\section{Introduction}

Inflationary expansion of the Universe is now believed to have existed 
before the radiation dominated era in the early Universe \cite{uobs,planck15}.
Although a lot of inflation models have been proposed by now \cite{slowroll}, 
we do not know which model can describe this phenomenon correctly. 
The relation between inflation and particle physics is also unclear. 
However, recent results of the CMB observations seem to have ruled out 
many of them already \cite{planck15}. This has been done by comparing
both values of the scalar spectral index and the tensor-to-scalar 
ratio obtained from the CMB observations and predicted values for them 
by each model.

If we take a large inflaton scenario in the slow-roll inflation 
framework, trans-Planckian values are required for realization of
sufficient $e$-foldings. In that case, we cannot answer the question 
why higher non-renormalizable terms do not contribute
to the inflaton potential. If we include such terms in 
inflaton potential, slow-roll conditions cannot be satisfied.
However, such kind of difficulty can be reconciled by 
compactifying the inflaton trajectory into a winding trajectory 
in the higher dimensional fields space. In that case, sufficient 
$e$-foldings along the trajectory can be obtained even if each field is 
kept in sub-Planckian regions. That possibility was studied in \cite{Yang} 
by introducing an idea about a \textit{spiralized inflation} in 
two-dimensional field space as a possible solution of the 
multifield slow-roll inflation. Its realizations have been proposed 
in the several frameworks, such as in string models \cite{Spiral}, 
SUSY models \cite{susy}, axions-based models 
\cite{axion, Natural, AlignNI, real, Dante} 
and a complex scalar model which has similar features with axions-based 
models \cite{complex, mac}. The higher order corrections 
to the potential realizing the spiralized inflation are found 
to affect significantly the predicted tensor-to-scalar ratio 
without changing the spectral index substantially \cite{corr}.    

In this paper we study an inflation model based on large but 
sub-Planckian inflaton. The model considered here has been 
proposed in \cite{bks}.
It could have an intimate connection to neutrino mass generation.
Although the required number for $e$-foldings is found to be realized 
in this model, the predicted tensor-to-scalar ratio by the model 
is too large compared with a central value of the up-dated observational 
results \cite{planck15,bkp}. Our main purpose is to study whether 
the favorable values for them can be obtained in this model. We also discuss
a possible connection to a certain particle physics model.        

\section{A sub-Planckian inflaton model} 
\begin{figure}[t]
\begin{center}
       \includegraphics[width=9cm]{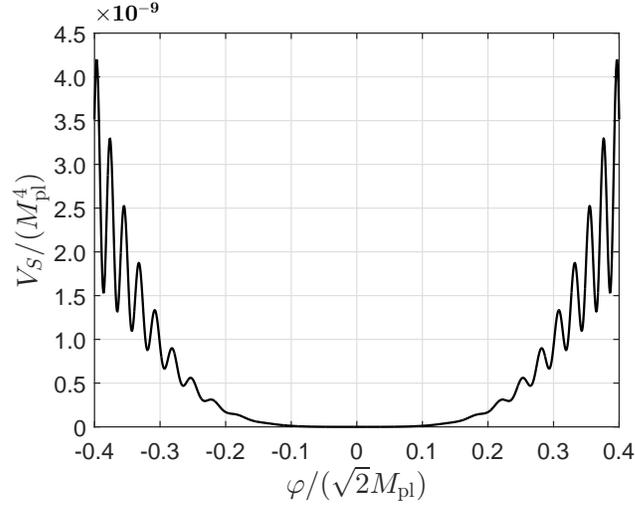}
\end{center}
\vspace*{5mm}

\caption{ Potential $V_S$ at a fixed $\theta$ 
for $n=2, m=1$ case. Other parameters are fixed 
at $c_1=1.1791 \times 10^{-7}$, 
$c_2 = 1.4$ and $\Lambda=0.05 M_{\rm pl}$.}
\label{figpot}
\end{figure}

The model studied here is defined by a complex scalar $S$ 
which has $Z_2$ odd parity. Its $Z_2$ invariant potential is assumed to 
be given such as \cite{bks}   
\begin{equation}
V_S = c_1\frac{(S^\dagger S)^n}{M_{\rm pl}^{2n-4}}
\left[1+ c_2\left\{ \left(\frac{S}{M_{\rm pl}}\right)^{2m} 
\exp\left(i\frac{S^\dagger S}{\Lambda^2}\right)
+ \left(\frac{S^\dagger}{M_{\rm pl}}\right)^{2m}
\exp\left(-i\frac{S^\dagger S}{\Lambda^2}\right)
\right\} \right]. \nonumber \\
\label{model1}
\end{equation}
If we use the polar coordinate $S=\frac{\varphi}{\sqrt 2}e^{i\theta}$,
this potential can be written as
\begin{equation}
V_{S}=c_1\frac{\varphi^{2n}}{2^nM_{\rm pl}^{2n-4}}\left[
1+ 2c_2\left(\frac{\varphi}{\sqrt 2 M_{\rm pl}}\right)^{2m}
\cos\left(\frac{\varphi^2}{2\Lambda^2}+2m\theta\right)\right],  
\label{infpot}
\end{equation}
where $c_2$ is assumed to satisfy 
$c_2< 0.5 \left(\frac{\varphi}{\sqrt 2M_{\rm pl}}\right)^{-2m}$.
As shown in Fig.~\ref{figpot}, 
$V_{S}$ has local minima with a potential barrier
$V_b\simeq \frac{c_1c_2\varphi^{2(n+m)}}{2^{n+m-2}M_{\rm pl}^{2(n+m-2)}}$
in the radial direction. These minima form a spiral-like valley
whose slope in the angular direction could be extremely small.
As a result, if we use the field evolution along this valley, 
slow-roll inflation is expected to be caused even for sub-Planckian 
values of $\varphi$ \cite{bks,mac}.  

\begin{figure}[t]
        \begin{center}
        \begin{subfigure}[b]{0.46\textwidth}
                \includegraphics[width=\textwidth]{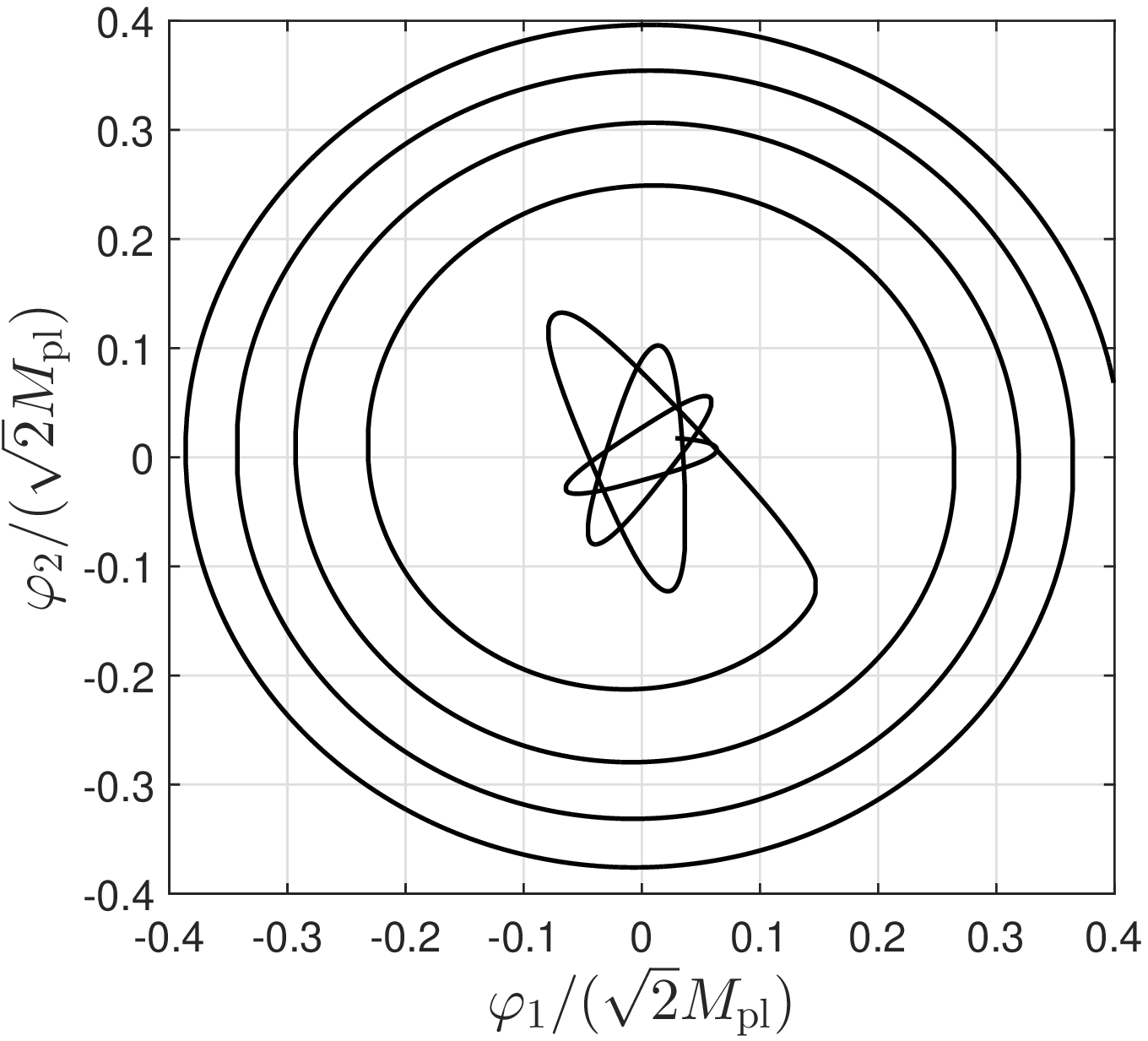}
                \caption{}
                \label{fig:dyn}
        \end{subfigure}%
    ~ %add desired spacing between images, e. g. ~, \quad, \qquad, \hfill etc.
      %(or a blank line to force the subfigure onto a new line)
        \begin{subfigure}[b]{0.55\textwidth}
                \includegraphics[width=\textwidth]{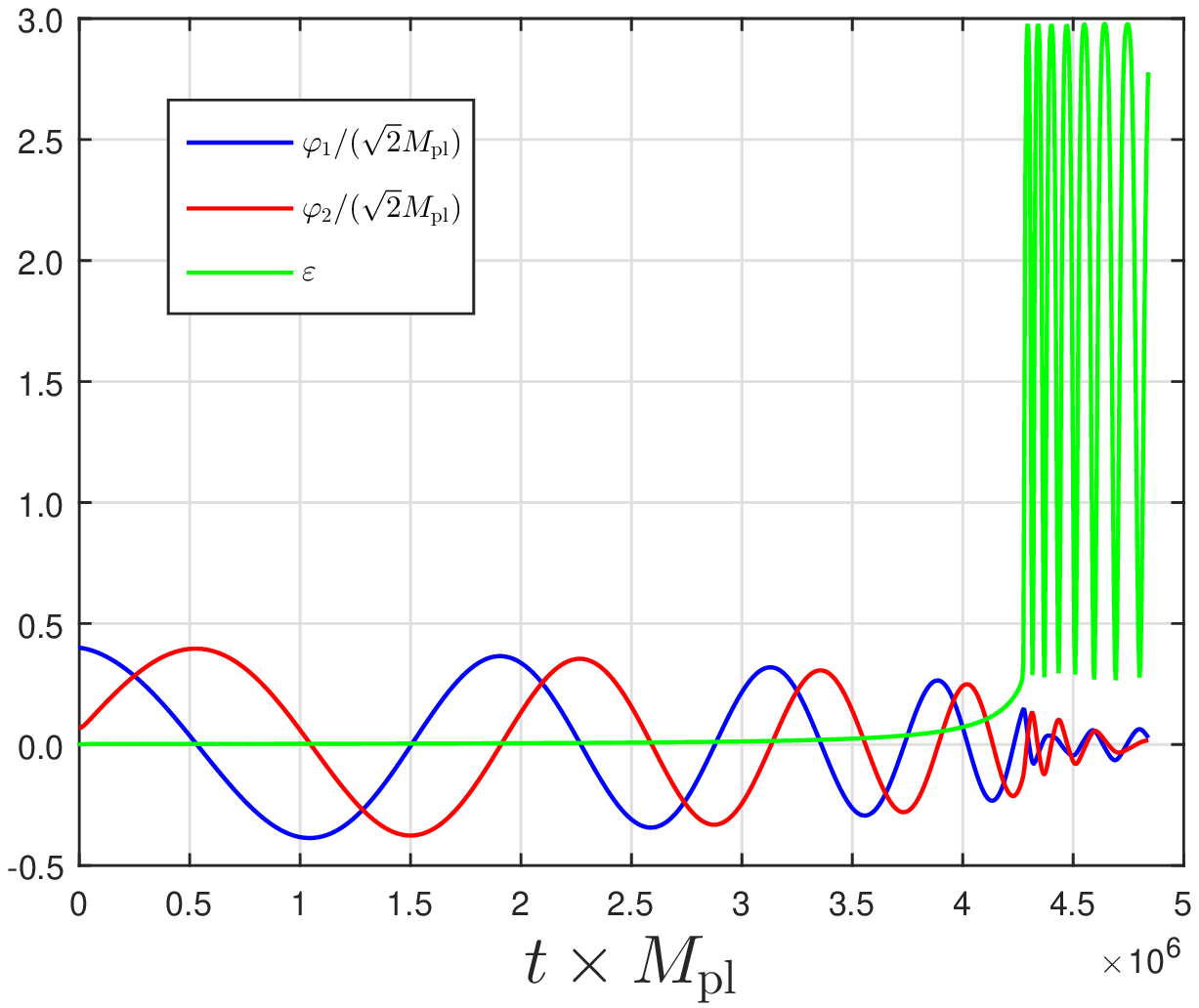}
                \caption{}
                \label{fig:evo}
        \end{subfigure}
				\end{center}
    ~ %add desired spacing between images, e. g. ~, \quad, \qquad, \hfill etc.
      %(or a blank line to force the subfigure onto a new line)
      %\caption{Pictures of animals}\label{fig:animals}
\caption{  Inflaton evolution for $n=2, m=1$ case. 
Parameters in the potential are fixed at the same values as
the ones used in Fig.~\ref{figpot}. Inflaton is assumed to be 
at a potential minimum initially.  In this case, 
it is numerically proven that the end of 
single field inflation
signed as turning point in the panel (a) is mostly realized much before 
$\varepsilon \simeq 1$, such that it is illustrated in the panel (b).  }
\label{fig:1}
\end{figure}

The evolution of the scalar field 
$S=\frac{1}{\sqrt{2}} (\varphi_1 + i \varphi_2)$ in this potential 
is dictated by the following equation of motion:
\begin{align}
\ddot{\varphi_i} + 3 H \dot{\varphi_i} = 
    -\frac{\partial V_S}{\partial \varphi_i} \qquad (i=1,2),
\label{evol}
\end{align}
where the Hubble parameter $H$ of the system is now written as 
$H^2 =\frac{1}{3M_{\rm pl}^2} \left(\sum_i \frac{1}{2} \dot{\varphi_i}^2 
+V_S \right)$ and $\frac{\partial V_S}{\partial \varphi_i}$ 
denotes partial derivative of the potential $V_S$ in the direction of 
field component $\varphi_i$. 
Taking $m=1$, the terms $\frac{\partial V_S }{\partial \varphi_i}$  
could be simply written for any number of $n$ as follows,
\begin{align}
\frac{\partial V_S}{\partial \varphi_1} = 
\frac{c_1 \left(S^\dagger S\right)^n}{M_{\rm pl}^{2n-4}} 
\left[ \frac{n\varphi_1}{\left(S^\dagger S\right)} \right.
&+ \left. \frac{c_2 \varphi_1}{M_{\rm pl}^{2}} 
\left\{ \frac{ n \left(\varphi_1^2 -\varphi_2^2\right)}
{\left(S^\dagger S\right)} + 2 
- \frac{2\varphi_1 \varphi_2}{\Lambda^2} \right\}  
\cos\left(\frac{ S^\dagger S}{\Lambda^2}\right) \right. \nonumber \\
&- \left. \frac{c_2 \varphi_1}{M_{\rm pl}^{2}} 
\left\{ \frac{ 2n\varphi_1 \varphi_2 }{\left(S^\dagger S\right)} 
+ 2 \frac{\varphi_2}{\varphi_1} 
+ \frac{ \left(\varphi_1^2 -\varphi_2^2\right)}{\Lambda^2} \right\} 
 \sin\left(\frac{ S^\dagger S}{\Lambda^2}\right) \right], 	
\end{align}
\begin{align}
\frac{\partial V_S}{\partial \varphi_2} = 
\frac{c_1 \left(S^\dagger S\right)^n}{M_{\rm pl}^{2n-4}} 
\left[ \frac{n\varphi_2}{\left(S^\dagger S\right)} \right.
&+ \left. \frac{c_2 \varphi_2}{M_{\rm pl}^{2}} 
\left\{ \frac{ n \left(\varphi_1^2 -\varphi_2^2\right)}
{\left(S^\dagger S\right)} - 2 
- \frac{2\varphi_1 \varphi_2}{\Lambda^2} \right\}  
\cos\left(\frac{ S^\dagger S}{\Lambda^2}\right) \right. \nonumber \\
&- \left. \frac{c_2 \varphi_2}{M_{\rm pl}^{2}} 
\left\{ \frac{ 2n\varphi_1 \varphi_2 }{\left(S^\dagger S\right)} 
+ 2 \frac{\varphi_1}{\varphi_2} 
+ \frac{ \left(\varphi_1^2 -\varphi_2^2\right)}{\Lambda^2} \right\}  
\sin\left(\frac{ S^\dagger S}{\Lambda^2}\right) \right]. 	
\end{align}
We may solve eq.~(\ref{evol}) numerically to see 
the evolution of $\varphi_i$.  
The initial value of each component $\varphi_i$ cannot be selected 
arbitrarily since the slow-roll behavior could be ruined depending on it.
If the initial position of the inflaton is located at a point higher 
than its next potential barrier, the inflaton could cross over 
it without realizing the slow-roll motion along the angular direction. 
The most simple setting for the initial value to realize the slow-roll 
is to take it at a potential minimum. 
An example of the evolution of the scalar field components 
is illustrated in Fig. \ref{fig:1}.

Now we describe features of the inflation induced by this field 
evolution in detail.  
The radial component $\varphi$ is assumed to take a large initial 
value on a local minimum in the radial direction. 
Even if $\varphi$ is not just on this minimum initially, 
it converges to a minimum point within a certain period of time 
as long as it starts to roll from the neighborhood of 
a minimum point of the valley initially. 
In that case, as shown in \cite{bks},
the model could cause sufficient $e$-foldings through the inflaton 
evolution along this spiral-like valley even for sub-Planckian 
values of $\varphi$. 
An inflaton field $\chi$ could be identified with
\begin{equation}
\chi\equiv a_e+\frac{\varphi_{e}^3}{6m\Lambda^2}-a=
\frac{\varphi^3}{6m\Lambda^2},
\end{equation}
where the subscript $e$ of the fields stands for the field value 
at the end of single field inflation.
The field $a$ is defined by using $\varphi$ as
\begin{equation}
da=\left[\varphi^2+\left(\frac{d\varphi}{d\theta}
\right)^2\right]^{1/2}d\theta
=\left[1 + 4m^2\left(\frac{\Lambda}{\varphi}\right)^4\right]^{1/2}
\varphi d\theta.
\label{infl}
\end{equation}
The number of e-foldings caused by $\chi$ during its slow-roll 
is given as
\begin{equation}
N=-\frac{1}{M_{\rm pl}^2}\int_\chi^{\chi_e} d\chi ~\frac{V_{S}}{V_{S}^\prime}
\equiv N(\chi)-N(\chi_e), 
\label{efold0}
\end{equation}
where $V_{S}^\prime=\frac{dV_{S}}{d\chi}$ and $N(\chi)$ is represented 
by using the hypergeometric function $F$ as
\begin{eqnarray}
N(\chi)&=&\frac{1}{6m^2n}
\left(\frac{M_{\rm pl}}{\Lambda}\right)^4
\left(\frac{\varphi}{\sqrt 2M_{\rm pl}}\right)^6\left[~1 
+\frac{6c_2m}{n(3+m)}\left(\frac{\varphi}{\sqrt 2M_{\rm pl}}\right)^{2m}
\right. \nonumber \\
&&\hspace*{2cm}\left. 
\times F\left(1,~\frac{3}{m}+1,~\frac{3}{m}+2,~2c_2\left(1+\frac{m}{n}\right)
\left(\frac{\varphi}{\sqrt 2M_{\rm pl}}\right)^{2m}\right)\right].
\label{efold}
\end{eqnarray} 

Slow-roll parameters 
$\varepsilon\equiv \frac{M_{\rm pl}^2}{2}
\left(\frac{V_{S}^{\prime}}{V_{S}}\right)^2$ and 
$\eta\equiv M_{\rm pl}^2\left(\frac{V_{S}^{\prime\prime}}{V_{S}}\right)$ 
for single field inflation can be represented by 
using the model parameters as
\begin{eqnarray}
&&\varepsilon=m^2\left(\frac{\sqrt 2M_{\rm pl}}{\varphi}\right)^6
\left(\frac{\Lambda}{M_{\rm pl}}\right)^4\left[
\frac{n-2c_2(m+n)\left(\frac{\varphi}{\sqrt 2M_{\rm pl}}\right)^{2m}}
{1-2c_2\left(\frac{\varphi}{\sqrt 2M_{\rm pl}}\right)^{2m}}\right]^2, \nonumber\\
&&\eta=m^2\left(\frac{\sqrt 2M_{\rm pl}}{\varphi}\right)^6
\left(\frac{\Lambda}{M_{\rm pl}}\right)^4
\frac{n(2n-3)-2c_2(m+n)(2m+2n-3)\left(\frac{\varphi}{\sqrt 2M_{\rm pl}}
\right)^{2m}}
{1-2c_2\left(\frac{\varphi_1}{\sqrt 2M_{\rm pl}}\right)^{2m}}.  
\label{slow}
\end{eqnarray}
If the $c_2$ term is neglected in these formulas, we can find very 
simple formulas for these slow-roll parameters at the time 
characterized by the inflaton value $\chi_\ast$.
They can be represented by using the $e$-foldings $N_\ast$, 
which is defined for $N(\chi_\ast)$ in eq.~(\ref{efold0}), as 
\begin{equation}
\varepsilon\simeq\frac{n}{6(N_\ast+N(\chi_e))}, \qquad 
\eta\simeq\frac{2n-3}{6(N_\ast+N(\chi_e))}.
\end{equation}
Thus, the scalar index $n_s$ and the tensor-to-scalar ratio
$r$ can be derived as \cite{bks}
\begin{equation}
n_s=1-6\varepsilon+2\eta\simeq 1-\frac{n+3}{3(N_\ast+N(\chi_e))}, \qquad 
r=16\varepsilon\simeq \frac{8n}{3(N_\ast+N(\chi_e))}.
\end{equation}

In order to see the features of this model, it may be useful to 
compare the model with single field inflation with monomial 
potential $\varphi^{\bar n}$. Since the $e$-foldings in this model
is written as $N\simeq \frac{1}{2\bar n}\frac{\varphi^2}{M_{\rm pl}^2}$,
the sufficient $e$-foldings require a trans-Planckian value for the 
inflaton $\varphi$. We note that the contribution of $\varphi_e$ 
is negligible in this case.
If we use this number of $e$-foldings $N_\ast\simeq N(\varphi_\ast)$, 
the slow roll parameters are expressed as 
\begin{equation}
\varepsilon=\frac{\bar n}{4N_\ast}, \qquad \eta=\frac{\bar n-1}{2N_\ast},
\end{equation}
and then $n_s$ and $r$ can be written as
\begin{equation}
n_s=1-\frac{\bar n+2}{2N_\ast}, \qquad r=\frac{4\bar n}{N_\ast}.
\end{equation}

It is easily found that $n_s$ and $r$ in both models 
have the same expression for $\bar n=\frac{2}{3}n$ in the limit $c_2=0$. 
However, we should remind that the present model works well 
only for the non-negligible $c_2$ since this term causes 
the potential barrier $V_b$ in the radial direction. 
$V_b$ makes the inflaton $a$ evolve along the 
spiral-like trajectory formed by the potential minima  
like a single field inflation.
This brings about a different feature for the model 
from the $\varphi^{\bar n}$ inflation scenario.

In this model,the single field inflation 
is expected to end at the time when 
$\frac{1}{2}\dot{\chi}^2\simeq V_b$ is realized.
If we apply the slow-roll approximation $3H\dot\chi=-V_{S}^\prime$
to the slow-roll parameter $\varepsilon$,
the inflation is found to end 
at $\varepsilon=\frac{3V_b\left(V_b +V_{S}\right)}{V_{S}^2}$.
Since $V_{S}>V_b$ is satisfied, the end of inflation could happen 
much before the time when  $\varepsilon\simeq 1$ is realized. 
In such a case, $N(\chi)\gg N(\chi_e)$ is not satisfied 
and then $N(\chi_e)$ has a substantial contribution to determine 
the number of $e$-foldings $N_\ast$ in eq.~(\ref{efold0}).
Thus, the smaller $N_\ast$ could be enough to
realize the same values for $n_s$ and $r$ in comparison with the 
$\varphi^{\bar n}$ inflation.\footnote{ 
Although this becomes clear especially 
in the small $c_2$ case, $\varphi_e$ could be well approximated 
as the $\varphi$ value at $\varepsilon=1$ in other cases.}.
We should also note that the values of $n_s$ and $r$ in this
model could deviate largely from ones predicted in the $\varphi^{\bar n}$ 
inflation model due to the non-negligible contribution from the 
$c_2$ term. 
Illustration given in Fig. \ref{fig:evo} justifies this argument 
numerically when whole contributions, including non-negligible $c_2$ term, 
are taken in to account. As it is expected, the end of inflation 
at which inflaton starts to oscillate around global minimum of the 
potential takes place much before $\varepsilon \simeq 1$.   
After this time, the inflaton falls in the reheating  process and 
produces lighter particles.

\section{Spectral index}
We estimate the scalar spectral index $n_s$ and the 
tensor-to-scalar ratio $r$ by taking account of the $c_2\not=0$ effect. 
Before it, we constrain parameters in the potential by using 
the normalization for the scalar perturbation found in the CMB.
The normalization for the scalar perturbation found in the CMB 
observations gives the constraint on the inflaton potential
$V_{S}$ at the time when the scale characterized 
by a certain wave number $k_\ast$ exits the horizon. 
The observation of CMB requires the spectrum of scalar perturbation
${\cal P}_{\mathcal{R}}(k)=A_s\left(\frac{k}{k_\ast}\right)^{n_s-1}$ to
take $A_s\simeq 2.43 \times 10^{-9}$ 
at $k_\ast=0.002~{\rm Mpc}^{-1}$ \cite{uobs}, 
$V_S$ should satisfy
\begin{equation}
\frac{V_{S}}{\varepsilon}=(0.0275 M_{\rm pl})^4,
\label{norm}
\end{equation}
where we use $A_s=\frac{V_S}{24\pi^2M_{\rm pl}^4\varepsilon}$.
If the $c_2$ term does not dominate the potential, we can represent 
the condition from this normalization constraint as
$c_1\simeq 9.5 \times 10^{-8}\frac{n}{N_\ast}
\left(\frac{\sqrt 2M_{\rm pl}}{\varphi_\ast}\right)^{2n}$
where $\varphi_\ast$ is a value of $\varphi$ at the time when 
the $e$-foldings is $N_\ast$. 
  
On the other hand, the $e$-foldings $N_\ast$ expected 
after the scale $k_\ast$ exits the horizon is dependent on the reheating 
phenomena and others in such a way as \cite{slowroll}
\begin{equation}
N_\ast\simeq 61.4-\ln\frac{k_\ast}{a_0H_0}
-\ln\frac{10^{16}~{\rm GeV}}{V_{k_\ast}^{1/4}}
+\ln\frac{V_{k_\ast}^{1/4}}{V_{\rm end}^{1/4}}-
\frac{1}{3}\ln\frac{V_{\rm end}^{1/4}}{\rho_{\rm reh}^{1/4}}.
\end{equation} 
This suggests that $N_\ast$ should be considered to have a value 
in the range 50 - 60.  
Taking these constraints into account, we estimate both $n_s$ and 
$r$ for the case where $N_\ast$ is in this range.

\begin{figure}[t]
\begin{center}
\begin{tabular}{|c|c|c|c|c|c|c|c|c|c|}\hline
$n$ &$c_1$ & $c_2$ & $\frac{\Lambda}{M_{\rm pl}}$ & 
$\frac{\varphi_1^\ast}{\sqrt 2 M_{\rm pl}}$ & $H_\ast$
& $N_\ast$ & $n_s$ & $r$ & $n_s^\prime$ \\ 
&   &&& &$(\times 10^{13}{\rm GeV})$&&&& \\ \hline 
\hline
%New data----------------------------------------------------------------
3 & 1.00 $\times 10^{-6}$ &1.5 & 0.05& 0.417& 6.528& 60.0&  0.967& 0.070& -0.00047 \\
%  & 8.4901 &1.5 & 0.05& 0.4061& 6.1745& 50.0&  0.9608& 0.0909& -0.00070 \\
  & 9.84 $\times 10^{-7}$ &1.7 & 0.05& 0.411& 5.914& 60.0&  0.964& 0.056& -0.00043 \\
%  & 8.4901 &1.7 & 0.05& 0.4016& 5.6444& 50.0&  0.9586& 0.0759& -0.00065 \\	
	& 8.62 $\times 10^{-7}$ &1.9 & 0.05& 0.406& 5.399& 60.0&  0.959& 0.040& 
-0.00032 \\
%  & 8.4901 &1.9 & 0.05& 0.3963& 5.1253& 50.0&  0.9542& 0.0596& -0.00056 \\	
%--------------------------------------------------------------------------
\hline
%New data----------------------------------------------------------------
2 &1.32 $\times 10^{-7}$& 1.1& 0.05& 0.394& 7.019& 60.0& 0.973& 0.058& -0.00043   \\
	&1.76 $\times 10^{-7}$& 1.1& 0.05& 0.384& 6.725& 50.0& 0.968& 0.072& -0.00061   \\
	&1.22 $\times 10^{-7}$& 1.6& 0.05& 0.383& 5.931& 60.0& 0.969& 0.039& -0.00040   \\
	&1.71 $\times 10^{-7}$& 1.6& 0.05& 0.374& 5.767& 50.0& 0.964& 0.052& -0.00059   \\
	&1.03 $\times 10^{-7}$& 1.9& 0.05& 0.374& 5.318& 60.0& 0.963& 0.026& -0.00035   \\
%	&1.5540 & 1.9& 0.05& 0.3668& 5.2156& 50.0& 0.9577& 0.0373& -0.00056   \\
%------------------------------------------------------------------------
\hline
%New data----------------------------------------------------------------
%1 &0.1215 & 0.5&0.05 & 0.3591& 5.1979& 60.0& 0.9788& 0.0338& -0.00033\\ 
1  & 1.36 $\times 10^{-8}$ & 0.5&0.05 & 0.349& 5.079& 50.0& 0.975& 0.041& -0.00046\\ 
  &7.45 $\times 10^{-9}$ & 1.6&0.05 & 0.333& 4.146& 60.0& 0.970& 0.015& -0.00036\\ 
  &1.02 $\times 10^{-9}$ & 1.6&0.05 & 0.326& 4.102& 50.0& 0.966& 0.019& -0.00052\\ 
  &6.15 $\times 10^{-9}$ & 1.8&0.05 & 0.327& 3.976& 60.0& 0.966& 0.011& -0.00035\\ 
  &8.77 $\times 10^{-9}$ & 1.8&0.05 & 0.320& 3.944& 50.0& 0.962&  0.016& -0.00052\\
%------------------------------------------------------------------------ 
\hline
\end{tabular}
\end{center}
\vspace*{2mm}

{\footnotesize Table~1\ Examples of the predicted values for the 
spectral index $n_s$ and the tensor-to-scalar ratio $r$ 
in this scenario with $m=1$.  }
\end{figure}
 
\begin{figure}[ht]
        \begin{center}
				      \begin{subfigure}[b]{0.33\textwidth}
                \includegraphics[width=\textwidth]{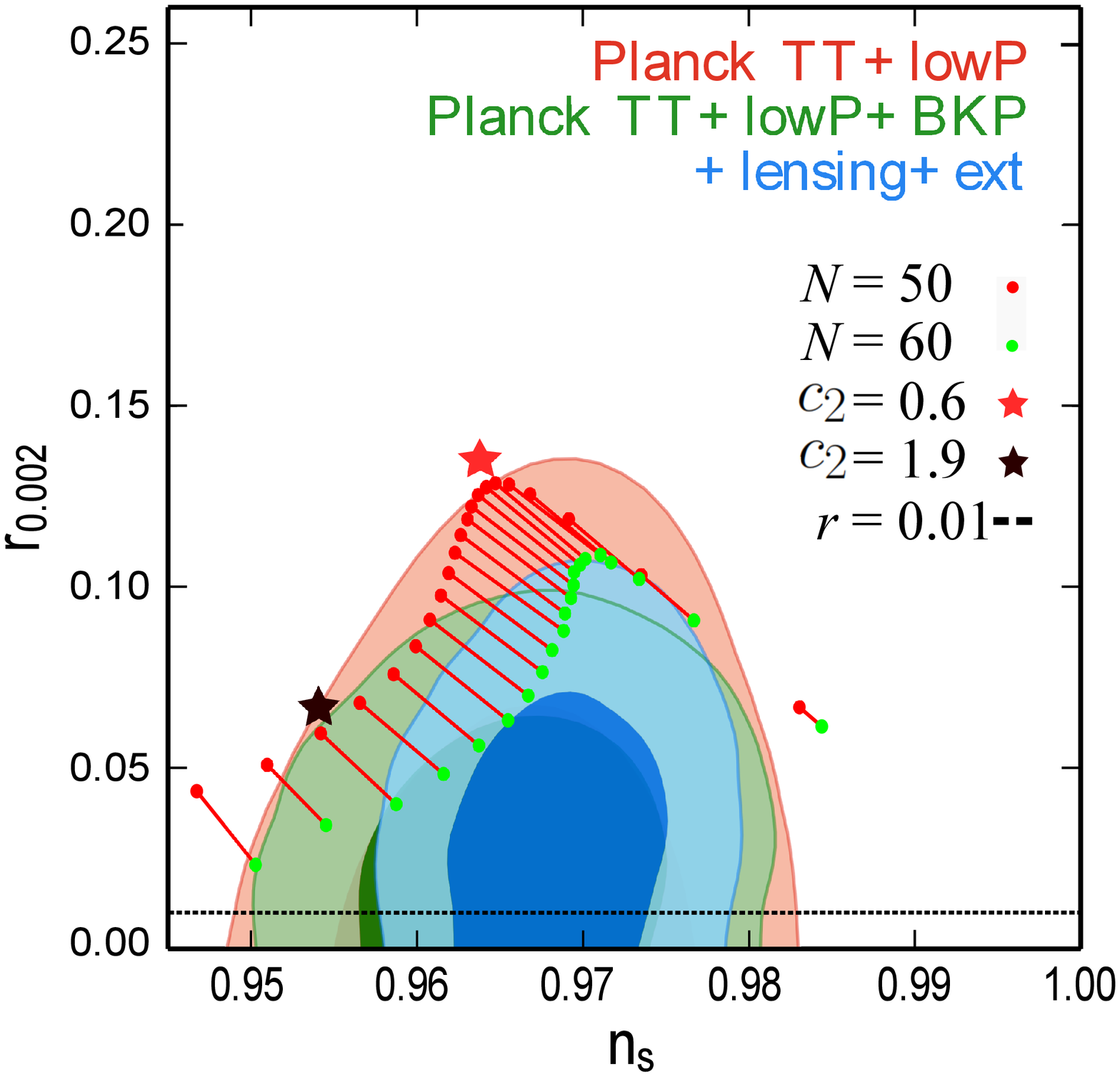}
                \caption{}
                \label{fig:n1a}
        \end{subfigure}%
				%-----
        \begin{subfigure}[b]{0.33\textwidth}
                \includegraphics[width=\textwidth]{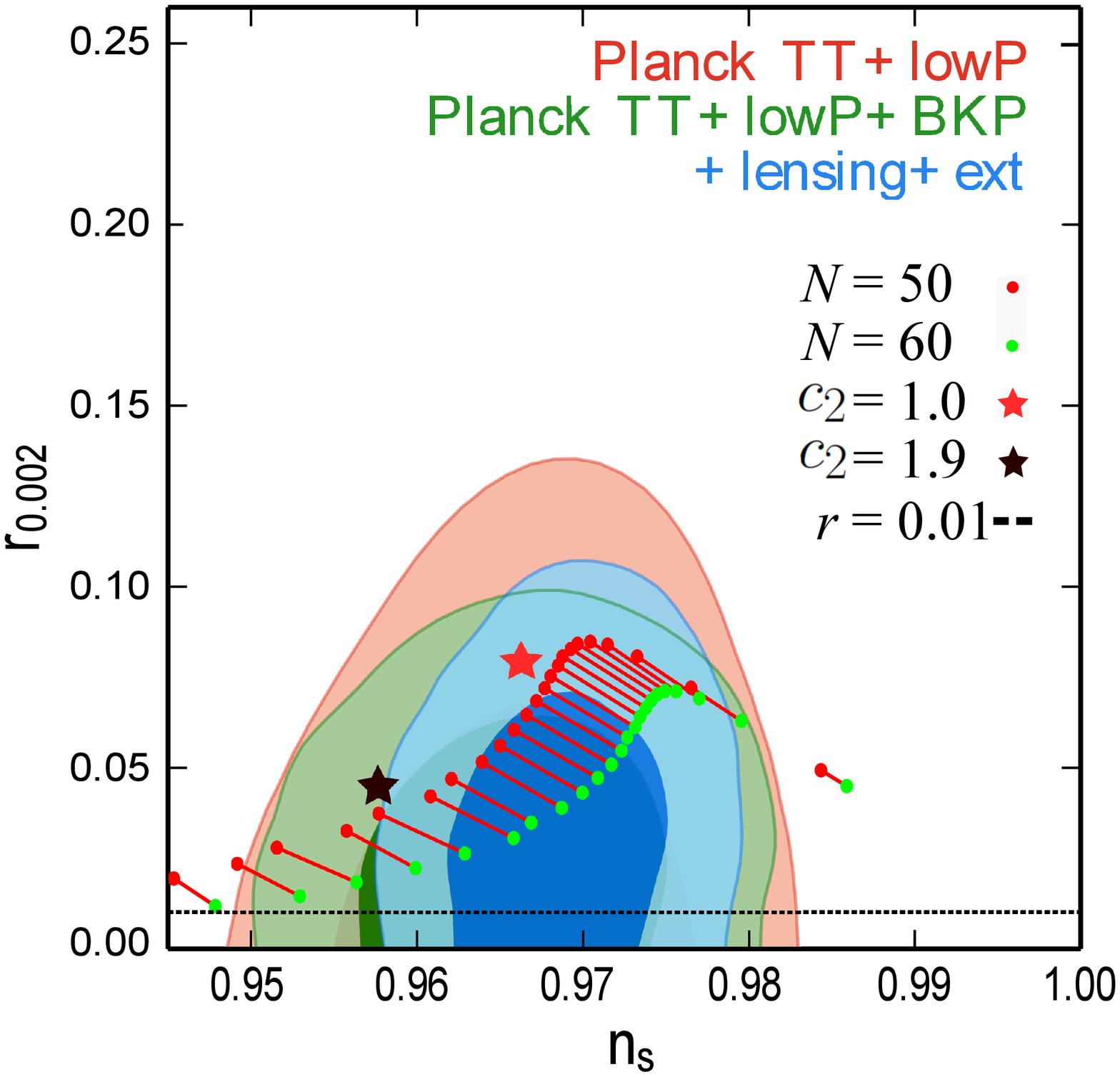}
                \caption{}
                \label{fig:n2}
        \end{subfigure}%
  ~ %add desired spacing between images, e. g. ~, \quad, \qquad, \hfill etc.
    %(or a blank line to force the subfigure onto a new line)
        \begin{subfigure}[b]{0.33\textwidth}
                \includegraphics[width=\textwidth]{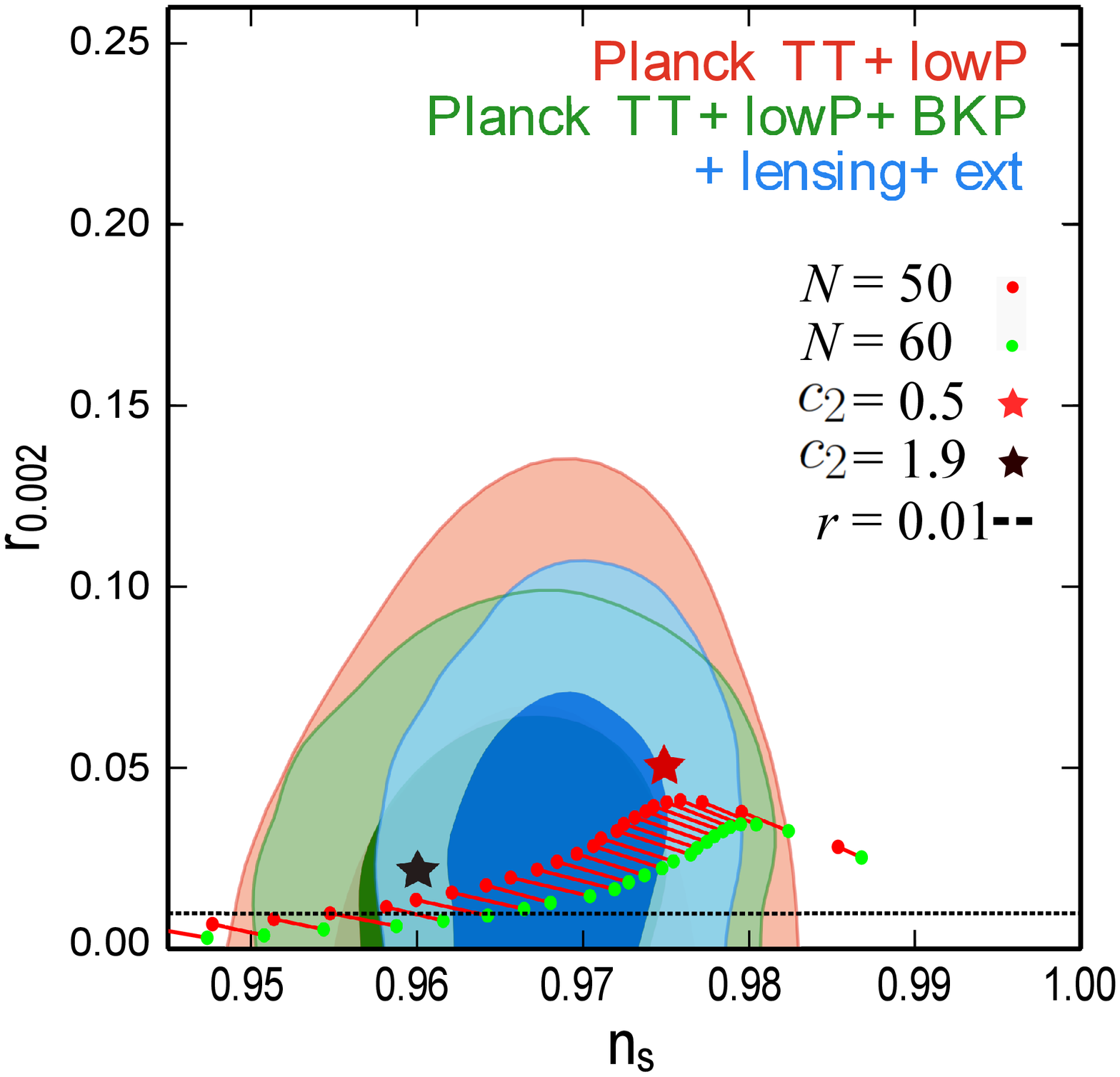}
                \caption{}
                \label{fig:n1}
        \end{subfigure}
				\end{center}
   ~ %add desired spacing between images, e. g. ~, \quad, \qquad, \hfill etc.
     %(or a blank line to force the subfigure onto a new line)
     %\caption{Pictures of animals}\label{fig:animals} 
\caption{~ Predicted regions in the $(n_s, r)$ plane 
are presented in panel (a) for $n=3$, 
in panel (b) for $n=2$, and in panel (c) for $n=1$. 
$\Lambda$ is fixed as $\Lambda = 0.05 M_{\text{pl}}$  
in all cases. The values of $c_1$ and $\varphi_\ast$ are given 
in Table 1 for representative values of $c_2$. 
Contours given in the right panel of Fig.~21 in Planck 2015 
results.XIII.\cite{planck15} are used here. Horizontal 
black lines $r=0.01$ represent a possible limit detected by LiteBIRD 
in near future. }
\label{fig:3}
\end{figure}

Numerical examples are shown in Table~1 for the cases $n=1,2,3$ with
a fixed $\Lambda$.\footnote{ We note that the first term of $V_S$ 
becomes $c_1M_{\rm pl}^2S^\dagger S$ and $c_1(S^\dagger S)^2$ for $n=1$ and 2, 
respectively.}
For given values of $c_2$, the values of $c_1$ and $\varphi_\ast$ are 
fixed so that the normalization condition given in eq.~(\ref{norm}) 
is satisfied and also $N_\ast$ takes its value in the imposed range 50 - 60.
Both $n_s$ and $r$ are estimated for them.\footnote{
If we apply the value of $A_s$ at 
$k_\ast=0.05$~Mpc$^{-1}$ \cite{planck15} 
to the present analysis using the same values of $c_{1,2}$ and 
$\Lambda$, $\varphi_\ast$ and $N_\ast$ are changed. 
This effect on $r$ is found to be $r_{0.05}\simeq 1.07r_{0.002}$ 
for the fixed values of $c_1$ and $c_2$ which give $n_s \simeq 0.971$ and
$r_{0.002} \simeq 0.1$ at $k_* = 0.002$ Mpc$^{-1}$. }
In  Fig.~\ref{fig:3}, 
we plot the predicted points in the $(n_s,r)$ plane by red and green 
circles, which correspond to $N_\ast=50$ and 60 respectively 
for every 0.1 of $c_2$ starting from $c_2=0.1$ on the right-hand side 
while $\Lambda$ is fixed as $\Lambda=0.05M_{\rm pl}$. 
We show the boundary values of $c_2$ by the red and black stars, 
for which either red or green circles are inside of the region of 
$2\sigma$ CL and $1\sigma$ CL of the latest Planck TT+lowP+ BKP+lensing+ext 
combined data for the $n=3$ and $n=1,2$ panels, respectively. 
They show that the present model with $c_2$ included in this interval 
are favored by the latest Planck data combined with others. 
The best fit result is obtained for the $n=1$ case.

As discussed above, the present model shows the similar behavior to
the monomial inflation models at least for the spectral index 
and the tensor-to-scalar ratio in the limiting case with the negligible 
$c_2$. However, if $c_2$ is not negligible, this feature could be changed
and these values largely deviate from the monomial inflation models.
Since the predicted region in the $(n_s,r)$ plane could be distinctive
from other inflation models, the model might be tested through 
future CMB observations. One of  the promising CMB 
observations would be LiteBIRD which is expected to detect the signal 
of the gravitational wave with $r > 0.01$ at more 
than $10 \sigma$ \cite{litebird}. 
Thus the whole of the predicted region could be verified in near future. 

Recent CMB results suggest that the running of the spectral 
index is consistent with zero at $1 \sigma$ level.
Thus, this can be an another useful test of the model.  
The running of the spectral index is known to be expressed 
by using the slow-roll parameters as
\begin{equation}
n_s^\prime\equiv \frac{dn_s}{d\ln k}
\simeq -24\varepsilon^2+16\varepsilon\eta-2\xi^2,
\end{equation}
where $\xi$ is defined as 
$\xi^2\equiv M_{\rm pl}^4\frac{V_S^\prime V_S^{\prime\prime\prime}}{V_S^2}$.
In the present model, it is written by using the model parameters as  
\begin{eqnarray}
\xi^2&=&2m^4\left(\frac{\sqrt 2M_{\rm pl}}{\varphi}\right)^{12}
\left(\frac{\Lambda}{M_{\rm pl}}\right)^8
\left[\frac{n-2c_2(n+m)\left(\frac{\varphi}{\sqrt 2M_{\rm pl}}\right)^{2m}}
{1-2c_2\left(\frac{\varphi}{\sqrt 2M_{\rm pl}}\right)^{2m}}\right. \nonumber \\
&&\left.\times \frac{n(n-3)(2n-3)-2c_2(n+m)(n+m-3)(2n+2m-3)
\left(\frac{\varphi}{\sqrt 2M_{\rm pl}}\right)^{2m}}
{1-2c_2\left(\frac{\varphi}{\sqrt 2M_{\rm pl}}\right)^{2m}}\right].
\end{eqnarray}
If we use the parameters given in Table~1, the running of the 
spectral index can be estimated in each case by using these formulas.
The results are shown in the last column of Table~1.
Although they are consistent with the latest Planck data,
they take very small negative values.
We might be able to use it for the verification of the model in future.

\section{Relation with particle physics}
Finally, we discuss the relation of the model with particle 
physics. Although we cannot clarify the origin of potential (\ref{model1})
at the present stage, we expect it might be produced through 
some non-perturbative effects of Planck scale physics.  
The complex scalar $S$ can play an important role in particle physics 
if we embed it in an extended standard model. 
As such an interesting example, we consider the radiative neutrino 
mass model proposed by Ma \cite{ma}. 
This model is given by the following Lagrangian for the neutrino sector: 
\begin{eqnarray}
-{\cal L}&=&\sum_{\alpha,k=1}^3\left(h_{\alpha k} \bar N_k\eta^\dagger\ell_\alpha
+h_{\alpha k}^\ast\bar\ell_\alpha\eta N_k+
\frac{M_k}{2}\bar N_kN_k^c 
+\frac{M_k}{2}\bar N_k^cN_k\right) \nonumber \\
&+&m_\phi^2\phi^\dagger\phi+m_\eta^2\eta^\dagger\eta+
\lambda_1(\phi^\dagger\phi)^2+\lambda_2(\eta^\dagger\eta)^2
+\lambda_3(\phi^\dagger\phi)(\eta^\dagger\eta)  
+\lambda_4(\eta^\dagger\phi)(\phi^\dagger\eta) \nonumber \\ 
&+&\frac{\lambda_5}{2}\left[ (\eta^\dagger\phi)^2+
(\phi^\dagger\eta)^2\right], 
\label{ma-model}
\end{eqnarray}
where $\ell_\alpha$ and $\phi$ are the doublet leptons and 
the ordinary doublet Higgs scalar in the standard model. 
Two types new fields are introduced in this model, that is, 
an inert double scalar $\eta$ and singlet fermions $N_k$. 
All their masses are assumed to be of $O(1)$~TeV.
New fields $\eta$ and $N_k$ are assigned odd parity of imposed $Z_2$ 
symmetry, although all the standard model contents have its even parity.
Since $\eta$ is assumed to have no vacuum expectation value, this $Z_2$
symmetry is exact and then neutrino masses cannot be generated at tree 
level. Neutrinos get masses through a one-loop diagram which has $\eta$ 
and $N_k$ in the internal lines as shown in the left-hand diagram of Fig.4. 
Moreover, the lightest neutral $Z_2$ odd field is stable
to be a good dark matter (DM) candidate. Thus, DM is an inevitable ingredient
for the neutrino mass generation in this model.  
The model has been clarified quantitatively to have interesting features 
through a lot of studies \cite{raddm,radlept}.

We can relate the present model to the Ma model by identifying 
the $Z_2$ symmetry in the present model with that in the Ma model. 
We assign its odd parity to the complex scalar $S$.    
If we take account of these symmetry, new terms which are subdominant
during the inflation period are introduced as invariant ones,
\begin{eqnarray}
-{\cal L}_S&=& \tilde m_{S}^2S^\dagger S+\frac{1}{2} m_{S}^2S^2+
\frac{1}{2} m_{S}^2S^{\dagger 2}+\kappa_1(S^\dagger S)^2 
+\kappa_2(S^\dagger S)(\phi^\dagger\phi)
+ \kappa_3(S^\dagger S)(\eta^\dagger\eta)
\nonumber \\
&-& \mu S\eta^\dagger\phi - \mu S^\dagger\phi^\dagger\eta.
\label{ext-model}
\end{eqnarray}
Here we note that the $\lambda_5$ term in eq.~(\ref{ma-model}) 
is also allowed under the imposed symmetry. 
However, since its $\beta$-function is proportional to itself
if an interaction $\mu S\eta^\dagger\phi$ in the last line of 
eq.~(\ref{ext-model}) is neglected, 
$\lambda_5=0$ is stable for radiative corrections.
On the other hand, if it is included in the Lagrangian, 
the $\lambda_5$ term can be induced through this interaction 
as the effective one at low energy regions after integrating out 
the heavy $S$ field. 

This can be easily seen through the neutrino mass generation. 
In the present extended model, the neutrino masses can be generated 
through the right-hand diagram of Fig.~4.
The neutrino masses obtained through this diagram 
can be described by the formula 
\begin{equation}
({\cal M}_\nu)_{\alpha\beta}=\sum_{k=1}^3\sum_{a=1,2}
\frac{h_{\alpha k}h_{\beta k}M_k\mu_a^2\langle\phi\rangle^2}{8\pi^2}
I(M_{\eta}, M_k, m_{\varphi_a}), 
\label{nmtr2}
\end{equation}
where $M_\eta^2=m_\eta^2+(\lambda_3+\lambda_4)\langle\phi\rangle^2$ and
$m_{\varphi_a}$ represents the mass of the real and imaginary component 
of $S$ which can be expressed as $m_{\varphi_1}^2=\tilde m_S^2+m_S^2$ and
$m_{\varphi_2}^2=\tilde m_S^2-m_S^2$. 
$\mu_a$ stands for $\mu_1=\frac{\mu}{\sqrt 2}$ and 
$\mu_2=\frac{i\mu}{\sqrt 2}$, respectively.
The function $I(m_a,m_b,m_c)$ is defined as
\begin{eqnarray}
I(m_a,m_b,m_c)&=&\frac{(m_a^4-m_b^2m_c^2)~\ln m_a^2}
{(m_b^2-m_a^2)^2(m_c^2-m_a^2)^2}+
\frac{m_b^2~\ln m_b^2}
{(m_c^2-m_b^2)(m_a^2-m_b^2)^2} \nonumber\\
&+&\frac{m_c^2~\ln m_c^2}
{(m_b^2-m_c^2)(m_a^2-m_c^2)^2}-
\frac{1}{(m_b^2-m_a^2)(m_c^2-m_a^2)}.
 \label{mnu2}
\end{eqnarray} 
If $m_{\varphi_a}^2 \gg M_k^2, M_\eta^2$ is satisfied and it corresponds to 
the present case, this formula is found to be reduced to
\begin{equation}
{\cal M}^\nu_{\alpha\beta}\simeq 
\left(\sum_{a=1,2}\frac{\mu_a^2}{m_{\varphi_a}^2}\right)
\sum_{k=1}^3\frac{h_{\alpha k}h_{\beta k}\langle\phi\rangle^2}{8\pi^2}
\frac{M_k}{M_\eta^2-M_k^2}\left[1 +\frac{M_k^2}{M_\eta^2-M_k^2}
\ln\frac{M_k^2}{M_\eta^2}\right],  
\end{equation}
which is equivalent to the neutrino mass formula obtained through
the left-hand diagram of Fig.~4 for the Ma model. This shows that 
$\lambda_5$ can be identified with $\sum_a\frac{\mu_a^2}{m_{\varphi_a}^2}$
as the effective coupling obtained at the low energy regions much 
smaller than $m_{\varphi_a}$.  
The key coupling for the neutrino mass generation in the Ma model
could be closely related to the inflaton interaction term in this 
extension. 

\begin{figure}[t]
        \begin{center}
                \includegraphics[width=15cm]{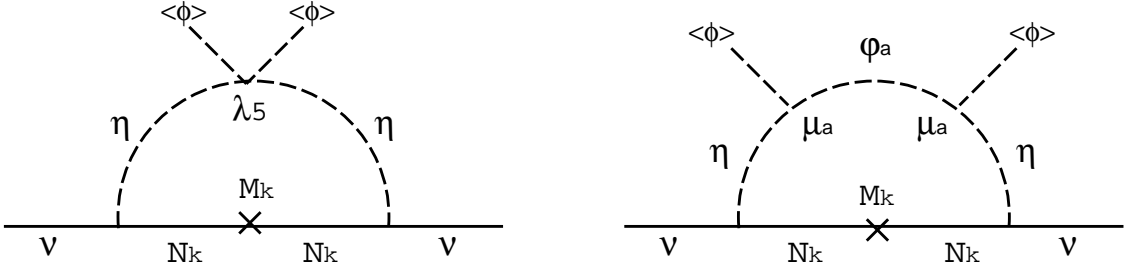}
         \end{center}
   \vspace*{5mm}
\caption{~One-loop diagrams contributing to the neutrino mass generation. 
The left-hand diagram is the one in the Ma model.
Lepton number is violated through the Majorama mass of $N_K$. 
The right-hand diagram is the one in the present extended model.
$\varphi_a$ represents the real and imaginary part of the singlet 
scalar $S$ defined by $S=\frac{1}{\sqrt 2}(\varphi_1+i\varphi_2)$.
$\mu_a$ is a dimensional coupling for $\varphi_a$ which is expressed 
as $\mu_1=\frac{\mu}{\sqrt 2}$ and $\mu_2=\frac{i\mu}{\sqrt 2}$. }
\label{fig:4}
\end{figure}

We should also note that the interesting feature for DM in the Ma model 
is completely kept in this extended model. 
We suppose that the $Z_2$ odd lightest field is the neutral real 
component of the inert doublet $\eta_R$. 
Its stability is guaranteed by the imposed $Z_2$ symmetry.
Since its relic abundance is determined by the coannihilations among the
components of $\eta$ which are controlled by the coupling constants 
$\lambda_{3,4}$ in eq.~(\ref{ma-model}), the results obtained 
in \cite{ham,ks} can be applied to the present model without affecting
the analysis in this paper.
They shows the required relic abundance $\Omega h^2=0.12$ could be
easily realized if either $\lambda_3$ or $\lambda_4$ takes 
a value of $O(1)$ for the $\eta_R$ with the mass of $O(1)$~TeV.
Thus, this extended model could give a simple explanation not only for 
the inflation but also for the neutrino masses and the DM abundance, 
simultaneously.   
 
\section{Summary}
We have considered an inflation scenario based on a complex 
singlet scalar. Special potential of this scalar constrains
the inflaton evolution along a spiral-like trajectory in the space 
of two degrees of freedom.
This makes the model behave like a single field inflation scenario. 
However, since the slop along this constrained direction is flat 
enough, inflaton can travel through trans-Planckian path.
As a result, the sufficient $e$-foldings can be realized even 
for sub-Planckian inflaton values. 
Serious potential problem in the large field inflaton could 
be solved in this model. 
Both the spectral index and the tensor-to-scalar ratio 
predicted in this model can be consistent with recently 
up-dated CMB observational results. 
Since these could take values in distinctive regions from other
inflation scenario, the model might be tested through future CMB 
observations. 

The inflaton in this model might be embedded into the extended 
standard model as an important ingredient. 
As such an example, we have discussed a possibility that the inflaton 
is an indispensable element in the radiative neutrino mass model,
where a certain quartic scalar coupling plays a crucial role in the neutrino 
mass generation. Since the inflaton causes this coupling 
as an effective one at low energy regions, it could have a close 
relation with particle physics in this extension.
The model might have another interesting feature. 
Reheating through the inflaton decay might give the origin of 
baryon number asymmetry through the generation of the lepton number 
asymmetry in a non-thermal way.  
Detailed study of this subject will be presented in future 
publication \cite{ks2}.
If it could be shown through explicit analysis, the problems in the standard 
model might be solved in a compact way in this extended model.
 
\section*{Acknowledgement}
R.~ H.~ S.~ Budhi is supported by the Directorate General of Higher 
Education (DGHE) of Indonesia (Grant Number 1245/E4.4/K/2012). 
S.~K. is supported by Grant-in-Aid for JSPS fellows (26$\cdot$5862).
D.~S. is supported by JSPS Grant-in-Aid for Scientific
Research (C) (Grant Number 24540263) and MEXT Grant-in-Aid 
for Scientific Research on Innovative Areas (Grant Number 26104009).

\end{document}